\documentclass[useAMS,usenatbib,usegraphicx,aas_macros]{mn2e}
\usepackage{graphicx, txfonts, color, soul,ulem, multirow, rotating}
\usepackage[bookmarks=true]{hyperref}
\usepackage{graphicx}
 
\usepackage[countmax]{subfloat}

\title[Exoplanet Transit Variability]{Exoplanet Transit Variability: Bow Shocks and Winds Around HD 189733b} 
\author[J. Llama et al.]{J. Llama$^1$\thanks{E-mail: joe.llama@st-andrews.ac.uk}, A.~A. Vidotto$^1$, M. Jardine$^1$, K. Wood$^1$, R. Fares$^1$  and T.~I. Gombosi$^2$\\
$^1$SUPA, School of Physics \& Astronomy, University of St Andrews, North Haugh, St Andrews,  KY16 9SS, UK\\ 
$^2$Department of Atmospheric, Oceanic and Space Sciences, University of Michigan, 1517 Space Research Building, Ann Arbor, MI 48109, USA
}

\begin{document}
 
\date{Accepted for publication in MNRAS}

\pagerange{\pageref{firstpage}--\pageref{lastpage}} \pubyear{2013}

\maketitle

\label{firstpage} 

\begin{abstract}
By analogy with the solar system, it is believed that stellar winds will form bow shocks around exoplanets.  For hot Jupiters the bow shock will not form directly between the planet and the star, causing an asymmetric distribution of mass around the exoplanet and hence an asymmetric transit. As the planet orbits thorough varying wind conditions, the strength and geometry of its bow shock will change, thus producing transits of varying shape.
We model this process using magnetic maps of HD 189733 taken one year apart, coupled with a 3D stellar wind model, to determine the local stellar wind conditions throughout the orbital path of the planet. We predict the time-varying geometry and density of the bow shock that forms around the magnetosphere of the planet and simulate transit light curves. 
Depending on the nature of the stellar magnetic field, and hence its wind, we find that both the transit duration and ingress time can vary when compared to optical light curves.
We conclude that consecutive near-UV transit light curves may vary significantly and can therefore provide an insight into the structure and evolution of the stellar wind.


 


\end{abstract}
\begin{keywords}
planets and satellites: general - planets and satellites: magnetic fields - planet-star interactions - planets and satellites: individual (HD 189733b) - stars: coronae - stars: individual (HD 189733) - stars: winds, outflows 
 \vspace{-0.25in}
\end{keywords}

\section{Introduction}
The discovery of hot Jupiters that orbit within fractions of an au to their host star are challenging our understanding of planet formation and evolution. Orbiting so close to their host star, such planets will be subjected to very different physical conditions to the gas giants in our own Solar System.

Star-planet interactions are thought to happen between hot Jupiters and their host star. Such interactions are believed to enhance the coronal activity of the host star \citep{Cuntz:2000ef}. One interaction between a hot Jupiter and its host star is thought to be magnetic, whereby reconnection occurs between the stellar and planetary magnetic fields. Tidal interactions which are a direct consequence of the planet orbiting so close to the parent star are also thought to occur. It has also been suggested that tidal interactions between the planet and star may serve to increase the magnetic activity of the star itself \citep{Cuntz:2000ef, Miller:2012gq}.  

For transiting exoplanets, asymmetries in the transit light curve provide a wealth of information about the planet and also the host star. Bumps in the transit light curve, caused by the planet occulting dark regions of the stellar disc provide an indirect method for inferring the location of star spots on the surface of the star \citep{SanchisOjeda:2011bwa, SanchisOjeda:2011hd}. On the Sun it is well known that the locations of Sunspots is cyclic and follows a so-called ``butterfly pattern". This cyclic nature of star spots is believed to be linked to the stellar dynamo \citep{Berdyugina:2005wb}. For other stars, continuous observations of transiting exoplanets provides us with the opportunity to reveal stellar butterfly patterns \citep{Llama:2012jl}.

 \begin{figure*} 
   \centering
   \includegraphics[width=6in]{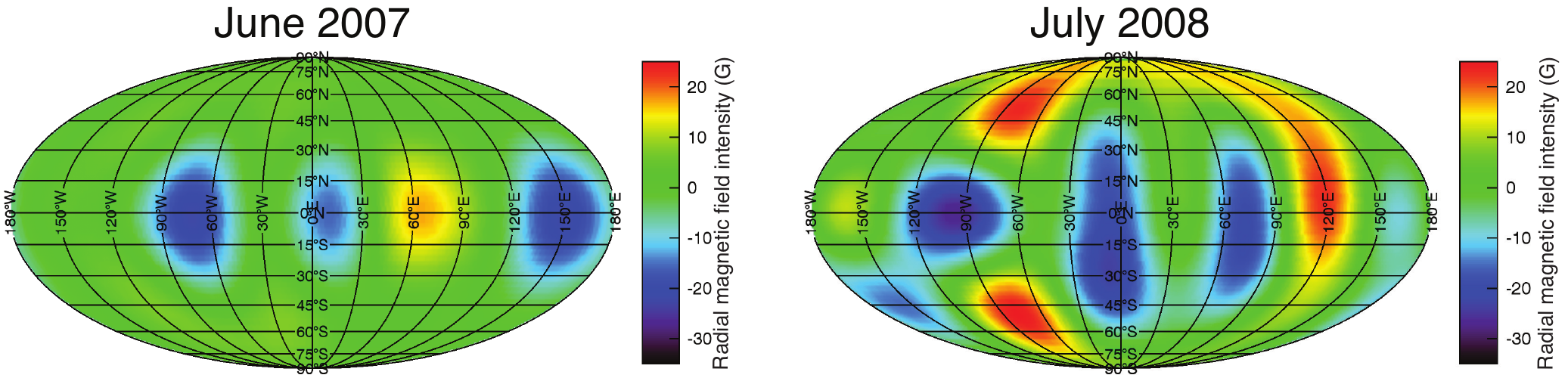} 
   \caption{Reconstructed surface magnetic maps of HD 189733 using Zeeman-Doppler Imaging from \citet{Fares:2010hq}. The left image is the magnetic map from June 2007 and the right image is from July 2008. In both cases, the maps show the distribution of radial magnetic field over the surface of the star. The radial field intensity is colour coded with blue being negative andred corresponding to positive field. In both cases the field strength varies from approximately $-40$G to $+25$G.  }
   \label{fig:zdi}
\end{figure*}

 
Timing asymmetries have also been observed in exoplanet transits. An early ingress in the light curve of WASP-12b observed by HST in the near-UV was first reported by \citet{Fossati:2010do} and later confirmed by \citet{Haswell:2012ft}. They found that not only did the near-UV transit begin before the optical transit, but also that the depth of the near-UV transit was much deeper. Such an asymmetry implies the presence of additional absorbing material in the exosphere of the planet that is non-uniformly distributed. There have been a number of solutions presented in the literature to explain such an asymmetry around the planet \citep{Lai:2010he, Vidotto:2010jha, Bisikalo:2013ev}.

One explanation is that very large, heavily irradiated exoplanets such as WASP-12b that orbit extremely close to their host star may fill and even overflow their planetary Roche lobe. Such an overflow would result in an accretion stream from the planet onto the star. If such an accretion stream were to be optically thick it could manifest itself as an asymmetry in the planets transit light curve \citep{Gu:2003hr,Ibgui:2010ev, Lai:2010he}.

Another possible explanation is that the planet may have a magnetic field that is interacting with the stellar wind, resulting in a bow shock forming around the magnetosphere of the planet \citep{Vidotto:2010jha}. Depending on the local conditions of the stellar wind, the bow shock may form ahead of the planet which would indeed cause an early ingress to be recorded in the light curve. \citet{Llama:2011de} modelled such a bow shock using a Monte-Carlo Radiative transfer code and found that they were able to fit the HST observations of WASP-12b, suggesting that the extra absorption and asymmetry could indeed be due to the presence of a bow shock surrounding the exoplanetary magnetic field. 

\citet{Vidotto:2010jha} also note that the presence of an early ingress in the light curve may therefore provide a method for detecting the presence of exoplanetary magnetic fields, the presence of which provides us with insight into the internal structure of the planet. For Earth sized planets, the presence of a magnetic field around the planet may be essential to sustain life.

  \begin{figure*} 
   \centering
   \includegraphics[width=6in]{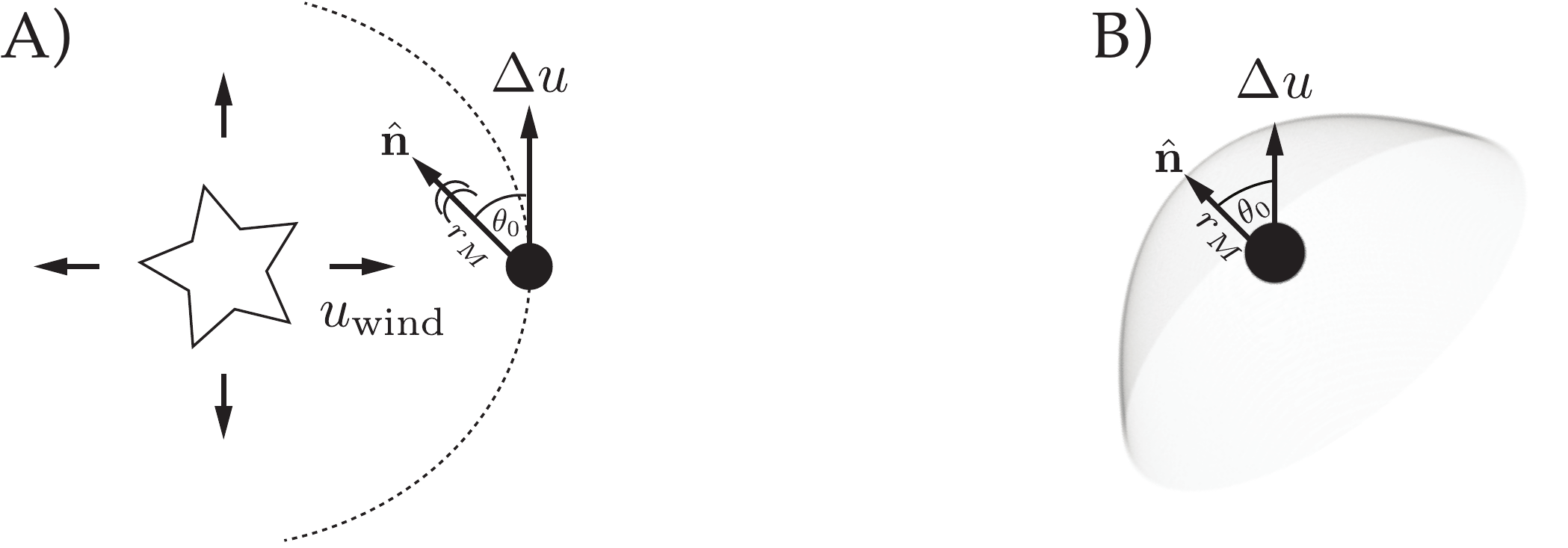} 
   \caption{Figure showing our shock model. A) shows the star-planet system and the orientation of the shock (adapted from \citet{Vidotto:2010jha}).  B) shows a zoomed-in region around the planet and shock.  $r_M$ is the distance to the nose of the shock. The angle $\theta_0$ is the angle made between the azimuthal direction of the planet and the vector $\hat{\textbf{n}}$, the normal of the shock nose. The planet is assumed to be spherical and completely dark. }
   \label{fig:cartoon}
\end{figure*}

One of the most extensively studied hot Jupiter hosting stars is the HD 189733 system. A bright K2V star with a mass and radius slightly smaller than that of the Sun it is an ideal target for investigating star-planet interactions and also for learning about the atmosphere of the planet. It was discovered to be hosting a hot Jupiter at an orbital distance of $8.8\,R_\star$ through transit observations by \citet{Bouchy:2005kv}. A list of parameters for the star, HD 189733, and the hot Jupiter, HD 189733b, are shown in Table \ref{tab:parameters}.

 \citet{Fares:2010hq} observed HD 189733 in June 2007 and again in July 2008, collecting spectra of the star that enabled them to recover the magnetic topology of the star. These spectra allowed them to look for enhanced activity due to star-planet interactions. They concluded no clear evidence of such interactions in the system during their epochs of observation. 
 
 Transits of HD 189733b have also been taken in wavelengths other than optical, looking for any direct evidence of an inflated atmosphere such as that of WASP-12b.  \citet{BenJaffel:2013ei} reported a potential early-ingress and increased absorption in the C II lines of HD 189733b. They carried out simulations of the planetary environment with a Parker model of the stellar wind and found that the stand-off distance between the front of the shock and the planet to be approximately $16.7\,R_P$ \citep{BenJaffel:2013ei, Parker:1958dn}.

 \citet{Bourrier:2013gl} have also observed HD 189733b in Lyman$-\alpha$ using HST/STIS in April 2010 and September 2011. In September 2011, they found an additional $14.4$ per cent absorption and hints of an early ingress which they accredit to atmospheric absorption; however, in April 2010 there was no presence of additional absorbing material or an early ingress. 
 

The presence of an early ingress and extra absorption in some but not all transit light curves of HD 189733b suggests that the presence of additional material in the exosphere of the planet may be transient and may depend on the local conditions of the stellar wind surrounding the planet.

 \citet{Vidotto:2011iw} discuss the potential variability in near-UV light curves due to the eccentricity of the planetary orbit, the presence of stellar magnetic cycles, and non-axisymmetric distributions of the stellar magnetic field. They reason that because the planet would interact with winds of varying conditions, the shock characteristic will vary throughout the orbit. As a consequence, the light curve will also vary as the absorption profile of the shocked material changes.


In this paper we explore the effects the stellar wind has on the conditions surrounding the transiting planet HD 189733b. By incorporating the reconstructed magnetic maps of \citet{Fares:2010hq}, acquired at two separate epochs, we are able to investigate both the spatial and temporal variations of the stellar wind. Since no simultaneous magnetic maps and transit observations of HD 189733 have been taken we do not attempt to reproduce any particular set of existing observations. Rather, in this work we follow the method outlined by  \cite{Llama:2011de} and simulate Mg transit observations that could have been observed by \textit{HST} during June 2007 and July 2008. 

If the presence of a bow shock is linked with the conditions of the stellar wind then we would expect the transit shape to vary with the magnetic evolution of the star. This coupling of magnetic maps and stellar wind simulations therefore enables us to illustrate how near-UV transits of HD 189733b would appear if the transit observations were taken simultaneously with spectro-polarimetric observations. We simulate near-UV light curves to match the observations of WASP-12b taken by \citet{Fossati:2010do}; however, our model is highly portable and could easily be modified to simulate other wavelengths.

\section{Methods}
\label{sec:methods}
\begin{table}
\centering
\caption{Fundamental parameters for the HD 189733 system (taken from \citet{2008ApJ...677.1324T}) and wind simulation assumptions.}
\label{tab:parameters}
\begin{tabular}{@{}lcr}
	\hline
	 & Parameter & Value \\
 \hline	
 	Stellar mass & $M_\star\,(\rm M_\odot)$ & 0.78 \\
	Stellar radius & $R_\star\,(R_\odot)$ & 0.76 \\	
	Stellar rotation period & $P_{\rm{rot}}\, (d)$ & 11.94\\
	Stellar metallicity & [M/H] (dex) & $-0.03$ \\
	Planetary orbital distance & $r_{\rm{orb}}\,(R_\star)$ & 8.8 \\
	Planetary orbital period & $T_{\rm orb}$ (days) & 2.2 \\
	Optical transit duration & $T_{\rm dur}$ (hours) & 1.827 \\
	Impact parameter & $b\,(R_\star)$ & 0.68 \\
	Planet radius & $R_p\,(R_J)$ & 1.138 \\ 
	Coronal base temperature & $T_0\,(K)$ & $2\times10^6$\\
	Coronal base density & $n_0\,($cm$^{-3})$ & $1\times10^9$\\
	Polytropic index & $\gamma$ & 1.1 \\
	Wind particle mean mass & $\mu\,(m_p)$ & 0.5\\
\hline
\end{tabular}
\end{table}

To investigate the shape and variability of near-UV transit light curves of the transiting planet HD 189733b we model the stellar wind and the resultant shock around the planet. In this work we improve on the model presented in \citet{Llama:2011de} which aimed to reproduce the near-UV observations of WASP-12b taken by \citet{Fossati:2010do}. In that work the shock was a spherical shell and the stellar wind model was a simple Parker solution \citep{Parker:1958dn}. Here we make use of a full three-dimensional MHD stellar wind model and we also use a more realistic shock model.  

\subsection{Stellar Surface Magnetic Maps}

\label{sec:zdi}
 \citet{Fares:2010hq} observed HD 189733 and re-constructed surface magnetic maps using the spectropolarimeters ESPaDOnS and NARVAL at the Canada-France-Hawaii Telescope at Pic du Midi. They reconstruct the surface magnetic field using Zeeman-Doppler Imaging (ZDI) (for a full description of ZDI see \citet{Donati:1997tc}). ZDI is a tomographic imaging technique that enables the large-scale magnetic field intensity and orientation to be reconstructed on the stellar surface using a series of circular polarised spectra of the star. The latest version of the ZDI code describes the magnetic field by its radial poloidal, non-radial poloidal and also toroidal components, all of which are expressed in terms of spherical harmonic expansions \citep{Donati:2006kl}.

The reconstructed magnetic maps of HD 189733 taken by \citet{Fares:2010hq} are shown in Figure \ref{fig:zdi}.  The left-panel shows the reconstructed magnetic map for June 2007 and the right-panel shows the map from July 2008. For June 2007, they find an average magnetic field strength of 22 G, whilst in July 2008 they find a slightly larger average field strength of 36 G \citep{Fares:2010hq}.

\begin{figure*} 
   \centering
   \includegraphics[width=6.8in]{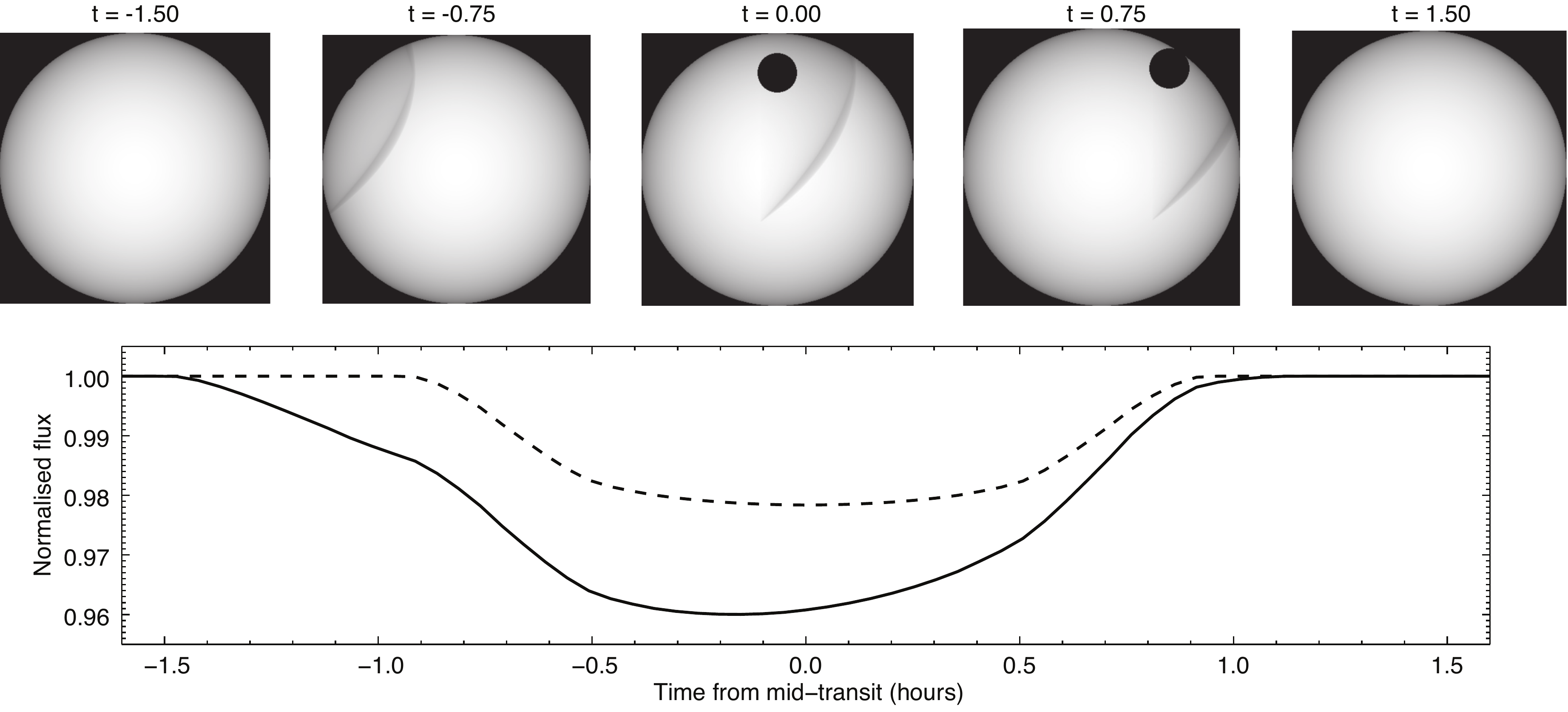} 
   \caption{A typical transit sequence of a planet and bow shock and resultant light curve for HD 189733 (top-row). The bottom graph shows the corresponding light curve (solid-line) with an optical transit (i.e. no bow shock detected) shown as a dashed line.  The first image shows the limb-darkened stellar disc before either the planet or the bow shock begin occulting the stellar disc. Because HD 189733b is a hot Jupiter, the shock begins transiting over the stellar disc before the planet. In this scenario, the shock blocks star light before the planet and so the near-UV transit event begins before the optical transit.  At mid-transit both the planet and bow shock are blocking star light, therefore the near-UV light curve has a deeper dip in flux than the optical light curve.  Because the shock is transiting ahead of the planet, it leaves the stellar disc before the planet resulting in the near-UV transit ending simultaneously with the optical light curve. The final image shows the end of the transit, once both the planet and shock have left the stellar disc. }
   \label{fig:transit images}
\end{figure*}

\subsection{Stellar Wind Model}
\label{sec:wind}
The magnetic surface maps are used as one of the boundary conditions in our stellar wind simulation. We use \texttt{BATS-R-US}, a three-dimensional mangetohydrodynamic (MHD) numerical code \citep{Powell:1999dq}. \texttt{BATS-R-US} solves the ideal MHD equations,
\begin{eqnarray}
	\frac{\partial\rho_w}{\partial t} + \nabla\cdot(\rho_w\textbf{u}_w) &= 0, \\
	\frac{\partial(\rho_w\textbf{u}_w)}{\partial t} + \nabla\cdot \left[\rho_w\textbf{u}_w\,\textbf{u}_w + \left(p_w + \frac{B_w^2}{8\pi}\right)\textbf{I} - \frac{\textbf{B}_w\,\textbf{B}_w}{4\pi}\right] &= \rho_w\textbf{g}, \\
	\frac{\partial\textbf{B}_w}{\partial t} + \nabla\cdot(\textbf{u}_w\,\textbf{B}_w - \textbf{B}_w\,\textbf{u}_w) &= 0, \\
	\frac{\partial\epsilon}{\partial t} + \nabla\cdot\left[\textbf{u}_w\left(\epsilon + p_w + \frac{B_w^2}{8\pi}\right) - \frac{(\textbf{u}_w\cdot\textbf{B}_w)\, \textbf{B}_w}{4\pi}\right] &=\rho\textbf{g}\cdot\textbf{u}_w,
\label{eqn:mhd}
\end{eqnarray}
for the wind mass density $\rho_w$, gas pressure $p_w$, the velocity of the plasma $\textbf{u}_w=(u_r,\, u_\theta,\, u_\varphi)$ and the magnetic field intensity $\textbf{B}_w=(B_r, \,B_\theta, \,B_\phi)$. The gravitational acceleration $\textbf{g}$ is caused by the star of radius $R_\star$ and mass $M_\star$. \texttt{BATS-R-US} assumes the gas to be ideal with $p=nk_BT$ where $T$ is the temperature and $n$ is the particle density of the stellar wind, i.e. $n=\rho/(\mu m_p)$ where $\mu m_p$ is the mean mass of the particle and $\gamma$ is the polytropic index (implying $p\propto\rho^\gamma$). Finally, the total energy density is expressed as 
\begin{equation}
	\epsilon = \frac{\rho u^2_w}{2} + \frac{p_w}{\gamma-1} + \frac{B_w^2}{8\pi}.
\label{eqn:energydensity}
\end{equation}
As a starting point for the simulation, the wind is assumed to be a thermally driven Parker wind \citep{Parker:1958dn}. At the stellar surface $(r=R_\star)$ we assume a coronal base temperature, $T_0 = 2\times10^6$ K, a base wind number density $n_0 = 10^9$cm$^{-3}$. This density value is chosen to reproduce the observed X-ray luminosity of HD 189733. The stellar parameters, $P_{\rm{rot}}$, $M_\star$ and $R_\star$ are given in Table \ref{tab:parameters}. These parameters provide the necessary information to solve the initial solution for the density, pressure and wind velocity. 

The final condition imposed on our initial set-up is that the magnetic field is assumed to be potential everywhere $(\nabla\times\textbf{B}_w=0)$. The initial solution to $\textbf{B}_w$ is found by imposing the observed ZDI maps at the surface of the star (see Section \ref{sec:zdi}) and by assuming the field to be entirely radial beyond a certain height above the star (known as the source-surface, assumed to be $4\,R_\star$). This approach is known as the potential field source surface method (e.g. \citet{Altschuler:1969cz, Jardine:2002ed}).
 
From the initial state of the simulation, the radial magnetic field distribution, $B_r$, along with the coronal base density and thermal pressure are kept the same at the stellar surface throughout the simulation. A zero radial gradient is then imposed to the remaining components of $\textbf{B}_w$ and $\textbf{u}_w=0$ in the frame co-rotating with the star. The outer boundaries at the edges of the grid have outflow conditions, i.e. a zero gradient is set to all the primary variables. The star is assumed to rotate as a solid body. 

Our grid is Cartesian and extends in $x$, $y$, and $z$ from $-20$ to $20~R_\star$, with the star placed at the origin of the grid. \texttt{BATS-R-US} uses block adaptive mesh refinement (AMR), which allows for variation in numerical resolution within the computational domain. The finest resolved cells are located close to the star (for $r \lesssim 2~R_\star$), where the linear size of the cubic cell is $0.01~R_\star$. The coarsest cell is about one order of magnitude larger (linear size of $0.3~R_\star$) and is located at the outer edges of the grid. The total number of cells in our simulations is around $80$ million. 

As the simulations evolve in time, both the stellar wind and magnetic field lines are allowed to interact with each other. The resultant solution, obtained self-consistently, is taken when the system settles into a steady state in the reference frame co-rotating with the star. This solution is able to diverge from the imposed initial potential solution and currents are allowed to form in the system. Also, the initially spherically symmetric hydrodynamical quantities $(\rho_w,\, p_w,\, \textbf{u}_w)$ are able to evolve into asymmetric distributions. The output from the wind simulation is a full three-dimensional grid that allows us to determine the local conditions experienced by the transiting planet HD 189733b. 

From the grid of densities we can compute the X-ray emission from the stellar corona. Assuming that the X-ray emission of the ionised coronal wind is caused by free-free radiation, we calculate the emissivity $\epsilon_\nu^{\rm ff}$ of an optically-thin, fully-ionised hydrogen plasma \citep{Rybicki:1986vo}
\begin{equation}\label{eq.xray}
\frac{\epsilon_\nu^{\rm ff}}{\rm{erg (s~cm^3 Hz)^{-1}}} = 1.7\times 10^{-38} \frac{n^2}{T^{1/2}}e^{-{h \nu}/{k_B T}} g_{\rm ff} \, ,
\end{equation}
where $\nu$ is the frequency of emission, $h$ the Planck's constant,  and $g_{\rm ff} \approx 1$ is the Gaunt factor \citep{Karzas:1962ji}. The X-ray luminosity $L_x$ is calculated by integrating  $\epsilon_\nu^{\rm ff}$ over X-ray frequencies and over the coronal volume. In practice, we compute $L_x$ in a cubed volume of $8R_\star\times8R_\star\times8R_\star$ with the star on its centre. Our results do not change significantly if we take a larger volume, as the bulk of the X-ray emission comes from small distances from the star. 

\subsection{Shock Model}

\begin{figure*} 
   \centering
   \includegraphics[width=6in]{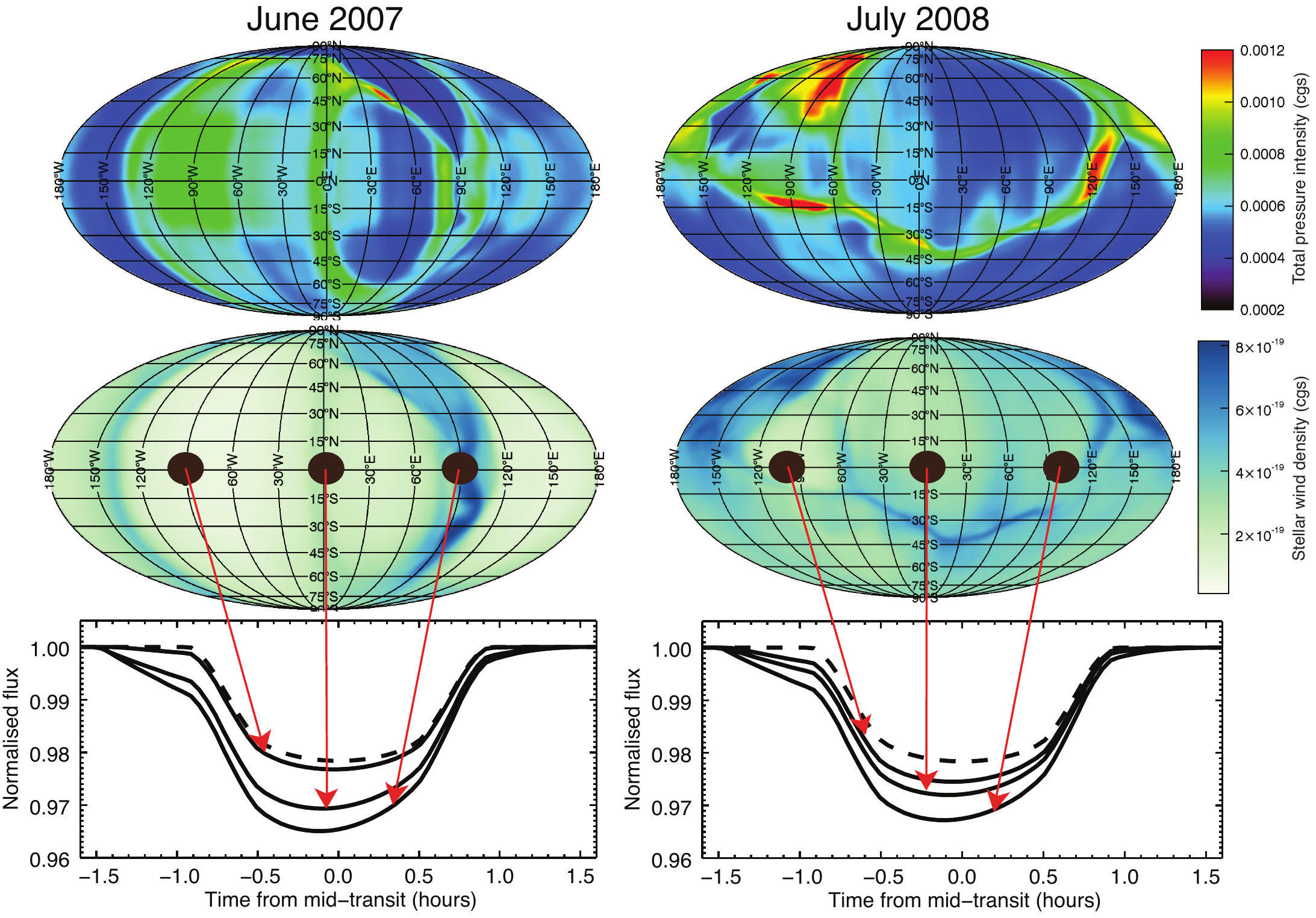} 
   \caption{Results from our simulations. The left-column shows the results from June 2007 and the right-column is for July 2008. The first row shows the total pressure (Equation \ref{eqn:ptot}) extracted from our wind simulations as a function of latitude and longitude at the orbital radius.  The middle row shows the density of the stellar wind at the orbital radius of the planet. Note that the planet orbits in the equatorial plane (lat $ = 0^\circ$). The bottom row shows simulated light curves of the planet and predicted bow shock (solid) and an optical light curve for reference (dashed). In both cases we have shown the deepest, shallowest and an intermediate light curve to highlight the expected variability. The arrows show the corresponding longitude for each transit.     }
   \label{fig:ptotplots}
\end{figure*}

The formation of a bow shock is a direct consequence of the interaction between the exoplanet's magnetic field and a super magneto sonic stellar wind. The distance from the planet at which the shock will form is directly related to the strength of the planetary magnetic field and also the strength of the stellar wind. The strength of the stellar wind will vary with location of the planet and also with the evolution of the stellar magnetic field. 

At a given distance, $r_M$ from the planet, pressure balance implies the total pressure from the stellar wind surrounding the planet balances the total pressure from the planetary material \citep{AAVidotto:2013hz}. We consider $r_M$ to characterise the size of the planet's magnetosphere. In reality, the distance to the bow shock is slightly larger than $r_M$.  For Earth, the location of the bow shock is a few tenths of $r_M$ further away from the planet. This force balance equation can be expressed as:
\begin{equation}
	\rho_w\Delta u_w^2 + \frac{B_w^2(r_{\rm orb})}{8\pi} + p_w = \frac{B_p^2(r_M)}{8\pi} + p_p,
\label{eqn:balance}
\end{equation}
where $|\Delta\textbf{u}_w|=|\textbf{u}_w-\textbf{u}_{\rm{Kep}}|$ is the relative velocity between the stellar wind and the planet's orbital motion. We assume the orbital motion of the planet to be Keplerian with
\[
u_{\rm{Kep}}=\left(\frac{GM_\star}{r_{\rm{orb}}}\right)^{1/2}.
\]
The values $\rho_w,\, B_w,\, \textbf{u}_w$ and $p_w$ are the local values of the stellar wind density, magnetic field strength and pressure at the location of the planet provided from the wind simulations (see Section \ref{sec:wind}) for all longitudes and latitudes. $r_{\rm{orb}}$ is the orbital radius of the planet from the host star.  $B_P(r_M)$ is the planetary magnetic field strength at the maximal extent of the planetary magnetosphere. To calculate an analytical expression for the value of $r_M$ we neglect the thermal pressures of the planet, $p_p$. We assume that the planet's magnetic field is dipolar in geometry, which implies that at the equatorial plane:
\begin{equation}
B_p\left(r_M\right) = \frac{B_p}{2}\left(\frac{R_p}{r_M}\right)^3,
\label{eqn:bplanet}
\end{equation}
where $B_p$ is the planet's magnetic field strength at the pole. For this investigation we assume the magnetic field strength to be comparable to that of Jupiter, i.e. $B_p=14$G. Substituting this expression for $B_p$ into Equation (\ref{eqn:balance}) we arrive at an expression for the maximal extent of the planetary magnetosphere: 
\begin{equation}
	\frac{r_M}{R_p} = \left(\frac{(B_p/2)^2}{8\pi(\rho_w\Delta u_w^2+p_w)+B_w^2}\right)^{1/6}.
\label{eqn:rm}
\end{equation}
The distance $r_M$ is depicted in Figure \ref{fig:cartoon}. The angle the shock normal makes with the azimthual direction of the planetary motion is defined as $\theta_0$ (see Figure \ref{fig:cartoon}). This angle is determined by the geometry of the stellar wind impacting on the planet. The equation for calculating the angle is given as 
\begin{equation}
	\theta_0 = \arctan\left(\frac{u_r}{|u_{\rm{Kep}}-u_\varphi|}\right),
\label{eqn:angle}
\end{equation}
where $u_r$ and $u_\varphi$ are the radial and azimuthal components of the stellar wind respectively \citep{Vidotto:2010jha}. If $\Delta \textbf{u}_w$ is mostly radial, then the wind particles will impact the planetary magnetosphere between the planet and the star and a so-called ``dayside shock" will form. Should the planet be orbiting much closer to the host star  then the flux of particles arriving at the planet will be ahead of the orbital motion of the planet resulting in an ``ahead-shock" forming ($\theta_0=0^\circ$). Shown in Figure \ref{fig:cartoon} is the intermediate case where the particles arriving at the planet form a bow shock with $0^\circ\leq\theta_0\leq90^\circ$.  

 \begin{figure*} 
   \centering
   \includegraphics[width=6in]{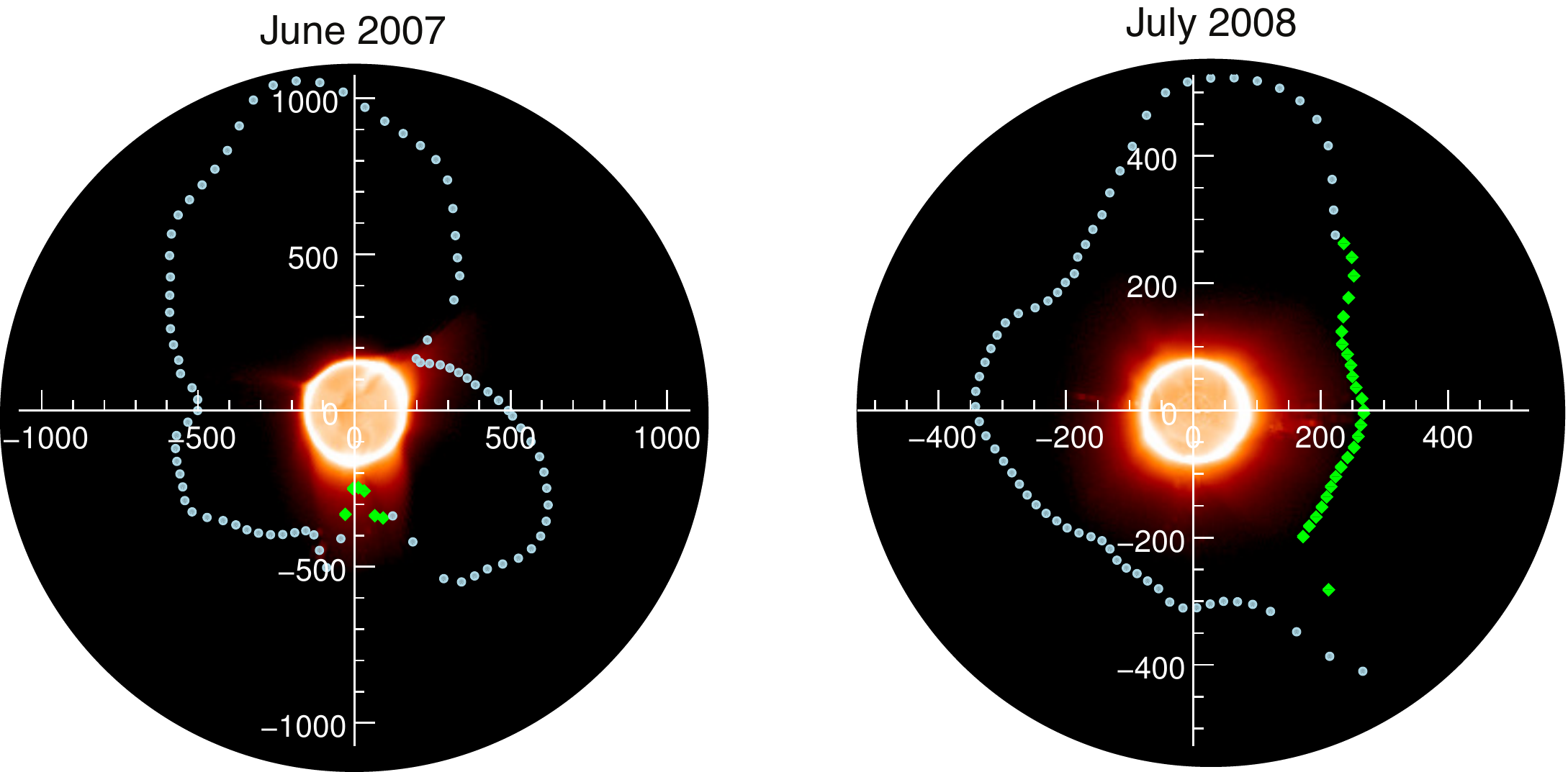} 
   \caption{Figure showing the view along the stellar rotation axis of the x-ray emission of the stellar corona for June 2007 (left) and July 2008 (right). Over plotted is the equatorial stellar wind velocity (in km s$^{-1}$) at the orbital radius of the planet. The blue circles denote an inward component to the magnetic field and the green diamonds denote an outward component.  }
   \label{fig:xray}
\end{figure*}

The distance to the nose of the shock, $r_M$, and the angle the shock makes with the orbital motion of the planet, $\theta_0$  define the shape of the resultant bow shock. The shock shape is prescribed by the model of \citet{Wilkin:1996bu}, where in two-dimensional polar coordinates $(\tilde{r},\tilde{\theta})$, centred on the planet, the distance from the planet to the shock can be expressed as: 
\begin{equation}
	\tilde{r}(r_m,\,\theta_0,\,\tilde{\theta}) = \frac{r_M}{\sin(\tilde{\theta}-\theta_0)}\sqrt{3\left(1-\frac{\tilde{\theta}-\theta_0}{\tan(\tilde{\theta}-\theta_0)}\right)}\,.
\label{eqn:shockshape}
\end{equation}

Figure \ref{fig:cartoon}B shows a typical shock shape. The shocked material is assumed to be adiabatic and so the density on the planet side of the shock is four times denser than the ambient wind density.  The ambient stellar wind density, $\rho_w$, is taken as an output from the wind simulation.  We begin by converting the mass density into a number density by assuming $n = \rho_w/\mu m_p$. Following \citet{Fossati:2010do}, we assume that in the near-UV C band. the dominant absorbing species in the shocked material is the Mg {\sc ii} doublet at 2795.5 $\AA$ and 2802.7 $\AA$. We calculate the number density of Mg {\sc ii} present by using the stellar metallicity, [M/H] (shown in Table \ref{tab:parameters}) and Equation 2 of \citep{Vidotto:2010jha}:
\begin{equation}
	\frac{n_{Mg}}{n_H} = 10^{(\epsilon_{Mg} -\epsilon_H)}10^{M/H},
\label{eqn:nmg}
\end{equation}
where, $\epsilon_{Mg}=7.53$ and $\epsilon_H=12.00$ are the solar abundances of Mg and H respectively. The absorption profile is computed by integrating the density of shocked material along the line-of-sight. The optical depth of the absorption profile is given by
\begin{equation}
	\tau = \int 4n_{Mg {\sc\rm II}}\sigma_{Mg {\rm\sc II}}\rm{d}S,
\label{eqn:tau}
\end{equation}
where $n_{Mg}$ {\sc ii} is the number of density of Mg {\sc ii} and $\sigma_{Mg \rm{{\sc II}}}$ is the extinction cross-section of Mg {\sc ii} and d$S$ is the path along the line-of-sight through the shocked material. The flux change, $F$, caused by the shocked material occulting the stellar disc is then calculated as:
\begin{equation}
	F = F_0 e^{-\tau},
\label{eqn:fluxchange}
\end{equation}
where $F_0$ is the un-occulted flux from region of the stellar disc being occulted by the shock.


\subsection{Transit Model}
\begin{figure*} 
   \centering
   \includegraphics[width=6in]{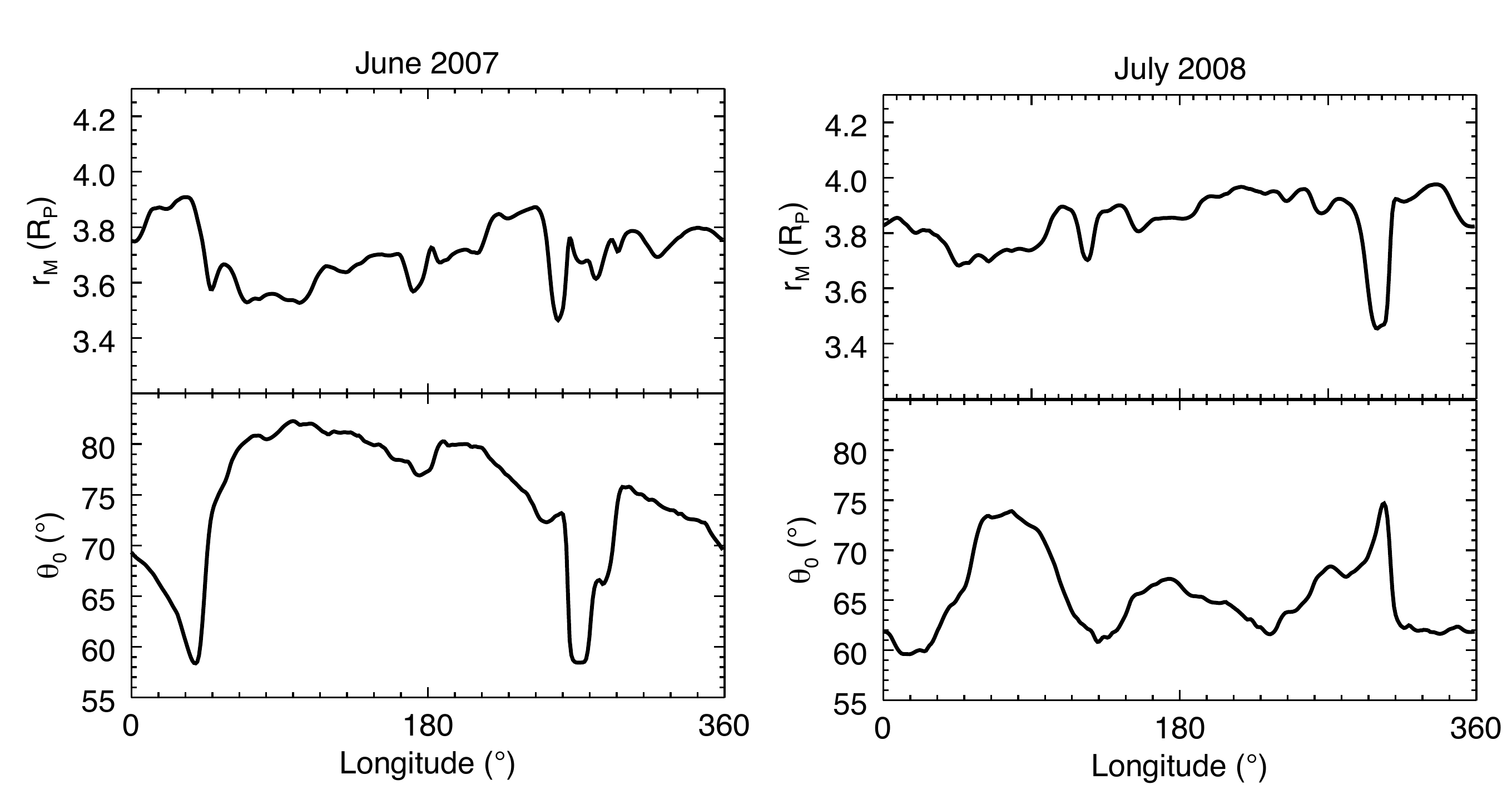} 
   \caption{Figure showing how the magnetosphere of the planet changes as the planet orbits around the star. The top row shows the variation in the size of the magnetosphere of the planet for June 2007 (left column) and July 2008 (right column). In both cases, the size of the magnetosphere, $r_M$ grows and shrinks approximately $15$ per cent as the planet orbits through various stellar wind conditions. The bottom row shows the variations in $\theta_0$, the angle the shock normal makes wit the azimuthal direction of the planetary motion. The orientation of the shock changes by up to $20^\circ$ as the planet orbits around the star.}
   \label{fig:planet plots}
\end{figure*}

For the HD 189733 system, $R_p/R_\star = 0.15463$ \citep{2008ApJ...677.1324T}. This ratio determines the dip in flux caused by the planet as it transits over the stellar disc and therefore sets the radius of the stellar disc in our light curve simulation. The stellar disc is limb-darkened using the limb-darkening law of  \citet{Claret:2004jy} where the intensity $I$ at any point on the stellar disc by the equation

\begin{equation} 
	\frac{I(\eta)}{I(0)} = 1-\sum^4_{j=1}a_j\left(1-\eta^{j/2}\right),
\label{eqn:fchange}
\end{equation}
where $I(0)$ is the emergent stellar intensity at disc centre and $\eta$ is the normalised radial distance into the stellar disc. The coefficients $a_j$ are chosen from \citet{Claret:2004jy} to match the stellar parameters and also the wavelength of the observations. Any radial distance larger than the radius of the star is set to zero to define the boundary of the stellar disc. Finally the intensity of the stellar disc is then normalised by the total intensity. The simulated limb-darkened disc of HD 189733 is shown in the top-left image of Figure \ref{fig:transit images}. 

We now simulate the transit of the planet and bow shock crossing the stellar disc. For this investigation, we assume the planet to be spherical and completely dark, i.e. any region of the stellar disc occulted by the planet will have an intensity of zero. The trajectory the planet takes over the stellar disc is determined by the impact parameter, $b$, which is a measure of the projected latitude of the stellar disc over which the planet appears to transit. For this investigation we adopt the impact parameter given by \citet{2008ApJ...677.1324T}, namely, $b=0.68$. The dashed line in Figure \ref{fig:transit images} shows a simulated optical transit light curve of HD 189733b.
  
At every point in the planet's orbit around the star we extract the local values of the stellar wind from our simulations. These values coupled with Equations (\ref{eqn:angle}) and (\ref{eqn:shockshape}) prescribe the shape and orientation of the bow shock. Equations (\ref{eqn:tau}) and (\ref{eqn:fchange}) are then used to calculate the optical depth of the absorption profile and resultant change in flux. The solid line in Figure \ref{fig:transit images}  shows an example transit of a planet and bow shock across the stellar disc. 

\section{Results}
\subsection{Stellar Wind}
The top and middle rows of Figure \ref{fig:ptotplots} are the results of our wind simulations for HD 189733 in June 2007 (left column) and July 2008 (right column). The images show the total pressure of the stellar wind (top row) and the wind density (middle row) at the orbital radius of the planet. The total pressure is the sum of the ram pressure, magnetic pressure and thermal pressure, i.e., 
\begin{equation}
	P_{\rm tot} = \rho_w\Delta u_w^2 + \frac{B_w^2}{8\pi} + p_w.
\label{eqn:ptot}
\end{equation}
Both the wind pressure and density are highly structured and vary with longitude and latitude. There is also very little similarity between the distribution of density and total pressure from June 2007 and July 2008 indicating the wind has evolved with the magnetic evolution of the star. The mass loss rate we find from our simulations is $\dot{\rm{M}} = 4.5\times10^{-13}\rm{M}_\odot/\rm{yr}$.




In our models, X-ray emission is modulated during the rotation of the star. This is a consequence of both the inhomogeneous distribution of surface magnetic fields and the inclination of the axis of rotation of the star towards the observer. For June 2007, the X-ray emission modulation can be as high as $18$~percent, while for July 2008, the modulation is not larger than $9$~percent.  This variability might have an impact on the detection of X-ray transits \citep{Poppenhaeger:2013wx}. 

Figure \ref{fig:xray} shows views along the stellar rotation axis the X-ray emission of the stellar corona. A polar plot of equatorial wind velocity at the orbital radius of the planet is over plotted. The blue values denote an inward radial component and green denote an outward component of the radial magnetic field, $B_r$. Regions of fast wind correspond to X-ray dark regions in the corona. Slow wind regions tend to lie over the largest helmet streamers in the coronal field. Therefore the structure that the stellar magnetic field imposes on the X-ray corona is correlated with the structure of the stellar wind.

\subsection{Planetary Magnetosphere Variability}
The variability in the stellar wind has a direct consequence on both the size of planetary magnetosphere and the orientation of the bow shock. This can easily be seen from Equations (\ref{eqn:rm}) and (\ref{eqn:angle}) which both depend on the properties of the stellar wind. Figure \ref{fig:planet plots} shows how $r_M$, the extent of the exoplanet's magnetosphere and the orientation of the bow shock, $\theta_0$, vary as the planet orbits around the star. Again, the left-column is the results from June 2007 and the right-column shows the results from July 2008. In both years, the size of the magnetosphere varies by $\sim 15$ per cent as the planet orbits around the star. We also find the value of $\theta_0$ to vary by $\sim 20$ per cent with a single orbit of the planet around the star.


\subsection{Light Curve Characteristics} 
A typical sequence of images showing a transit of a bow shock and planet over the stellar disc is shown in the top panel of Figure \ref{fig:transit images} and the corresponding light curve is shown underneath with an optical transit (i.e. no bow shock detected) shown as a dashed line. 

Because HD 189733b is a close-in hot Jupiter, we find $60^\circ\leq\theta_0\leq85^\circ$, which means that part of the bow shock transits ahead of the planet. This can clearly be seen in the second image of the top-panel of Figure \ref{fig:transit images} where the projected shock begins transiting over the stellar disc before the planet. As a consequence, the dip in light in the near-UV light curve occurs earlier than the dip in the optical transit.

\begin{figure*} 
   \centering
   \includegraphics[width=6in]{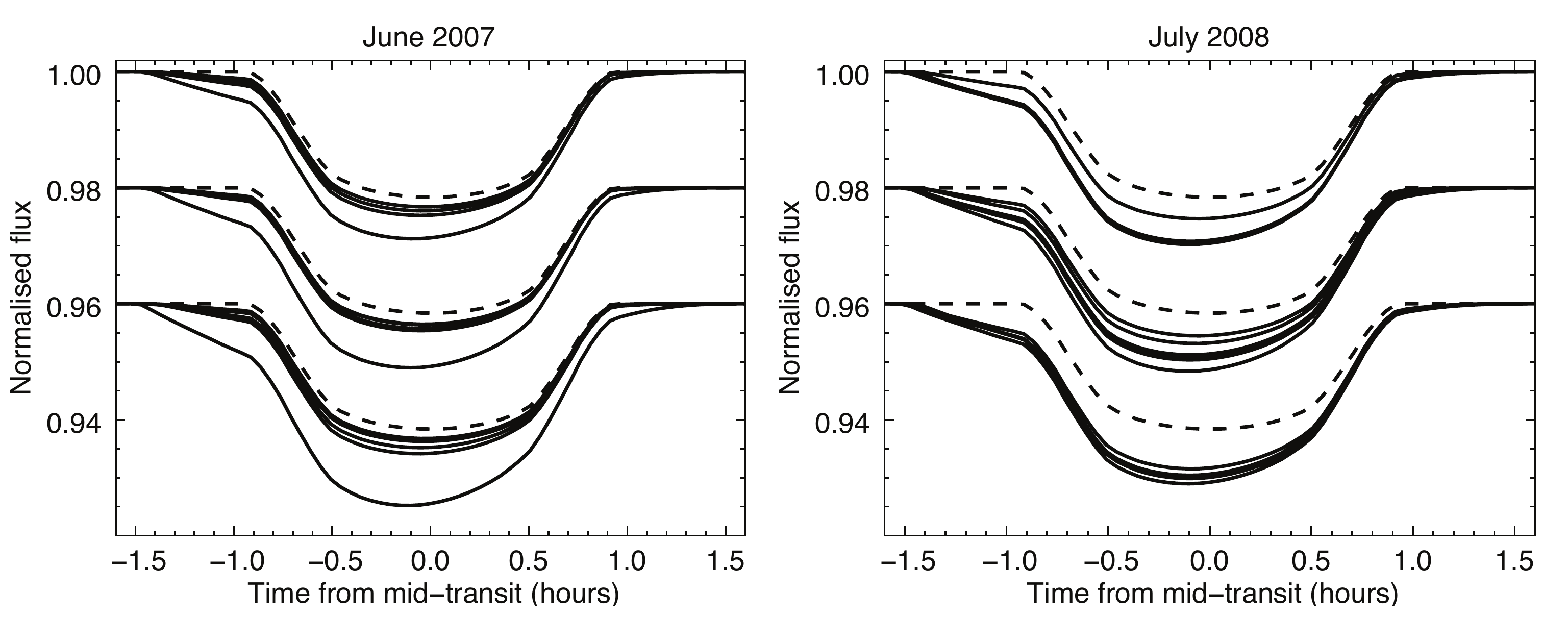} 
   \caption{Figure illustrating the expected variability with consecutive transits of HD 189733b. The left-panel is for June 2007 and the right-panel is for July 2008. Each time the planet transits the star will have rotated $\sim70^\circ$ and so the local wind conditions will be different to the previous transit. As a consequence the resultant near-UV light curve is also expected to be different. In each panel we show the expected light curves for three rotations of the star (each rotation separated by an offset in flux). The near-UV light curves are shown as solid lines and the optical as a dashed line. For June 2007 we find that only one of the transits during each rotation is significantly deeper and earlier when compared to the optical transit and so may be missed by observations. For July 2008 we find that the majority of the transits are both deeper and begin earlier than the optical light curve.   }
   \label{fig:ordered}
\end{figure*}

At mid-transit, both the shock and planet are occulting the stellar disc and so the depth of the light curve is deeper than that of the optical light curve. At the end of the transit, however, the shock leaves the stellar disc before the planet. This results in the transit event finishing simultaneously with the optical transit, creating an asymmetry in the near-UV light curve. 

It is worthy of note that the presence of an intermediate bow shock $(\theta_0\neq90^\circ)$ already breaks the symmetry of the light curve. If the shock were to be a purely ``dayside" shock $(\theta_0=90^\circ)$ then the light curve would be symmetric but could still be much deeper than its optical counterpart. 

\subsection{Light Curve Variations}
The changes in the shape and orientation of the bow shock coupled with the changes in wind density result in variations in observed transit light curves. The density of shocked material is directly related to the stellar wind density and so the detection of an early ingress and increased absorption requires the density of the wind to be high enough to cause a recordable dip in the light curve.

For each location of the planet we simulate a transit light curve for the planet and bow shock using the local values of wind density and pressure from our simulations. 
The bottom row of Figure \ref{fig:ptotplots} shows a selection of the simulated transit light curves for June 2007 and July 2008 (left and right panels respectively). In both cases, we show the deepest, shallowest and an intermediate simulated transit. The corresponding optical transits are shown as dashed lines. 

In the majority of cases, the near-UV light curves show more absorption than the optical transit. This is due to the presence of the bow shock in addition to the planet causing additional absorption of star light. All our simulated transits begin earlier than the optical transits, this is again because part of the shock begins transiting over the stellar disc before the planet.

The deepest transits occur when the density of the stellar wind is highest. In these cases, the early-ingress is clearly visible in the light curve. However, when the density of the stellar wind is lower, the shock is not dense enough to block enough  additional star light to cause a significant deviation from the optical light curve. In this case, the presence of a bow shock may go undetected. These results therefore indicate that the density of the shocked material and therefore the stellar wind play an important role in the detection of an early ingress and increased absorption. 

To quantify the variability of our light curves we calculate the range of values in two observables that would be measured through transit photometry. Firstly, we calculate the variability in the depth of the transit, $\Delta F$ which is the value of normalised flux at mid-transit.  Table \ref{tab:results} shows the variations in $\Delta F$ for both June 2007 and July 2008. 
  \begin{table}
 \centering
 \caption{Table showing two predicted observables from our simulations: The transit depth variations ($\Delta F$), and the timing variations between optical and near-UV transit ingresses ($\Delta t = t_{\rm op} - t_{\rm uv}$).}
 \label{tab:results}
 \begin{tabular}{llcc}
 \hline
  & & June 2007 & July 2008\\
  \hline
 \parbox[t]{2mm}{\multirow{3}{*}{\rotatebox[origin=c]{90}{$\Delta F$ }}} & Min &	0.977	&	0.975\\
 & Max &	0.965	&	0.967	\\
 & Avg &	0.975	&	0.971\\
  \hline
 \parbox[t]{2mm}{\multirow{3}{*}{\rotatebox[origin=c]{90}{$\Delta t$ (mins)}}} & Min &	0.00	&	3.65 \\
 & Max &	29.23  &	25.58\\
 & Avg &	4.85    &   14.62	\\ 
 \hline
 \end{tabular}
 \end{table}
We are also able to investigate the timing differences between the beginning of the optical transit and near-UV transit, $\Delta t = t_{\rm op} - t_{UV}$. Table \ref{tab:results} shows the variations in $\Delta t$ we find in our simulated light curves, which are extremely varied. In some cases the presence of the early ingress completely disappears whereas at other times it can also be as much as 30 minutes. It is worthy of note that the size of the planet's magnetosphere scales with the strength of the magnetic field, $B_p$ (Equation \ref{eqn:rm}). In this work we have assumed the planet to have a magnetic field strength similar to Jupiter (i.e. $B_p=14$G). If the value of $B_p$ were to change we would also expect the timing difference, $\Delta t$ to also change.

\section{discussion}

In this work we have shown that near-UV light curves of hot Jupiter's are expected to vary with the properties of the stellar wind. The presence of a magnetospheric bow shock causes not only the depth of the transit to change but also the timing of the transit ingress may be variable. 

We have assumed throughout this work that the shocked material is due to the presence of a magnetospheric bow shock. Heavily irradiated planets such as hot-Jupiter's are likely to exhibit significant mass loss. If the relative velocity of such outflows impacting with the stellar wind were to be supersonic, then they may also provide a mechanism for producing additional near-UV absorption.  This would likely require a planetary mass loss rate much larger than that estimated by \citet{Poppenhaeger:2013wx}. 

The star HD 189733 has a rotational period of 11.94 days and the planet has an orbital period of 2.2 days \citep{2008ApJ...677.1324T}. Therefore every time the planet transits over the stellar disc, the star will have rotated approximately $70^\circ$. From the two observed epochs of HD 189733 the likelihood of a transit observation being similar in consecutive near-UV transits is clearly dependent on the structure of the stellar wind. Figure \ref{fig:ptotplots} suggests that the structure of the stellar wind varies over approximately $~20^\circ-30^\circ$ in longitude. From this we would expect that consecutive near-UV transit light curves would be different. Figure \ref{fig:ordered} shows consecutive simulated light curves of HD 189733b for three rotations of the star for each of our simulations, i.e. June 2007 (left-column) and July 2008 (right-column). We have separated each rotation of the star by a flux value of $0.002$ for clarity. For each case we find that the consecutive recovered light curves vary in both transit depth and also starting time. For June 2007, we find that the likelihood of recovering a very deep, early transit to be quite unlikely; this is simply because of the relatively low degree of structure in the stellar wind. In July 2008 however, we find the consecutive light curves to have similar depths and starting time.  

We note that in \citet{Fossati:2010do} the error bars on their observations are of the order 1\% of the normalised flux. We would therefore not expect to be able to differentiate between the optical and near-UV light curves in all cases shown in Figure \ref{fig:ordered}. From this finding, one may conclude that the presence of a bow shock is time dependent; however, it may be the case that it is undetected. Indeed  \citet{LecavelierDesEtangs:2012jq} observed HD 189733b in Lyman$-\alpha$ at two different epochs and found no significant deviation from the optical transit in one epoch, whilst in the second they find a transit absorption depth of $14.4 \pm 3.6\%$.

In all cases, it is worth of note that the planetary magnetosphere is not changing significantly in size, rather the dominant factors in recovering an early ingress and additional absorption appears to be the density of the stellar wind and also the orientation of the shock normal $\theta_0$. As $\theta_0\rightarrow0^\circ$ (an ``ahead-shock") the path-length through the shock increases and so the early-ingress becomes more detectable.  

The size and orientation of the exoplanet's magnetosphere is of significant importance when we consider the atmosphere of the exoplanet. The magnetosphere shields the planet from the stellar wind and so should the magnetosphere be growing and shrinking significantly with time then this may expose regions of the planetary atmosphere to the stellar wind. The wind may then erode the atmosphere from the planet. Whilst this is not of great consequence for hot Jupiters, this may have severe consequences for smaller exoplanets that may sustain life \citep{AAVidotto:2013hz}.

 \section*{Acknowledgments}
JL acknowledges the support of an STFC studentship. AAV is supported by an RAS Fellowship. 
 
 \bibliography{hd189} 
\label{lastpage}

\end{document}